\documentclass[a4paper]{article}

\usepackage{INTERSPEECH2019}
\usepackage{cite}
\usepackage{color}
\usepackage{pifont}
\usepackage{soul}
\usepackage{booktabs}
\usepackage{multirow}
\usepackage{url}
\usepackage{times}
\usepackage[ruled,vlined,linesnumbered]{algorithm2e}

\title{Combining Adversarial Training and  Disentangled Speech Representation  for Robust Zero-Resource Subword Modeling}
\name{Siyuan Feng, Tan Lee, Zhiyuan Peng}
\address{
  Department of Electronic Engineering, The Chinese University of Hong Kong, Hong Kong}
\email{siyuanfeng@link.cuhk.edu.hk,  tanlee@ee.cuhk.edu.hk, jerrypeng1937@gmail.com}

\begin{document}
\maketitle
\begin{abstract}
This study addresses the problem of unsupervised subword unit discovery from untranscribed speech. It forms the basis of the ultimate goal of ZeroSpeech 2019, building text-to-speech  systems without  text labels. In this work,  unit discovery  is formulated  as a pipeline of phonetically discriminative feature learning and unit inference. One major difficulty in robust unsupervised feature learning is dealing with speaker variation. Here the robustness towards speaker variation is achieved by applying adversarial training and FHVAE based disentangled speech representation learning. A comparison of the two approaches as well as their combination is studied in a DNN-bottleneck feature (DNN-BNF) architecture. Experiments are conducted on ZeroSpeech 2019 and 2017. Experimental results on ZeroSpeech 2017 show that both approaches are effective while the latter is more prominent, and that their combination brings further marginal improvement in across-speaker condition. Results on ZeroSpeech 2019 show that in the ABX discriminability task, our approaches significantly outperform the official baseline, and are competitive to or even outperform the official topline. The proposed unit sequence smoothing algorithm improves synthesis quality, at a cost of slight decrease in ABX discriminability.

\end{abstract}

\section{Introduction}
\label{sec:intro}
Nowadays speech processing is dominated by deep learning techniques. Deep neural network (DNN) acoustic models (AMs) for the tasks of  automatic speech recognition (ASR) and speech synthesis have shown impressive performance for major languages such as English and Mandarin. Typically, training a DNN AM requires large amounts of transcribed data. For a large number of low-resource languages, for which very limited or no transcribed data are available, conventional methods of acoustic modeling are ineffective or even inapplicable.

In recent years, there has been an increasing research interest in zero-resource speech processing, i.e., only a limited amount of  raw speech data (e.g. hours or tens of hours)  are given while no text transcriptions or linguistic knowledge are available. The Zero Resource Speech Challenges (ZeroSpeech)   2015 \cite{versteegh2015zero}, 2017 \cite{dunbar2017zero} and 2019 \cite{dunbar2019zero} precisely focus on this area. One problem tackled by ZeroSpeech 2015 and 2017 is \textit{subword modeling}, learning frame-level speech representation that is discriminative to subword units and robust to linguistically-irrelevant factors such as speaker change. The latest challenge ZeroSpeech 2019 goes a step further by aiming at building text-to-speech (TTS) systems without any text labels (\textit{TTS without T}) or linguistic expertise. Specifically,  one is required to build an unsupervised subword modeling sub-system to automatically discover phoneme-like units in the concerned language, followed by applying the learned units altogether with speech data from which the units are inferred to train a TTS. Solving this problem may partially assist psycholinguists in understanding young children's language acquisition mechanism \cite{dunbar2019zero}.

This study addresses unsupervised subword modeling  in ZeroSpeech 2019, which is also referred to as acoustic unit discovery (AUD).
It is an essential problem and forms the basis of TTS without T.
The exact goal of this problem  is to represent  untranscribed speech utterances by discrete subword unit sequences, which is slightly different from subword modeling in the contexts of ZeroSpeech 2017 \& 2015.
In practice, it can be formulated as an extension to  the previous two challenges. For instance, after learning the subword discriminative feature representation at frame-level, the discrete unit sequences can be inferred by applying vector quantization methods  followed by collapsing  consecutive repetitive symbolic patterns. 
In the previous two challenges,  several unsupervised representation learning approaches were proposed for comparison, such as cluster posteriorgrams (PGs) \cite{chen2015parallel,ansari2017unsupervised,heck2017feature}, DNN bottleneck features \cite{shibata2017composite,chen2017multilingual}, autoencoders (AEs) \cite{renshaw2015comparison,kamper2015unsupervised}, variational AEs (VAEs) \cite{Feng2019improving,Chorowski2019unsup} and siamese networks \cite{thiolliere2015hybrid,Zeghidour+2016,Riad2018}.

One major difficulty  in unsupervised subword modeling is dealing with speaker variation. The huge  performance degradation caused by speaker variation reported in ZeroSpeech 2017 \cite{dunbar2017zero} implies that speaker-invariant  representation learning is crucial and remains to be solved. In ZeroSpeech 2019, speaker-independent subword unit inventory  is  highly desirable in building a TTS without T system. 
In the literature, many works focused on improving the robustness of unsupervised feature learning towards speaker variation.
One direction is to apply  linear transform methods. Heck et al. \cite{heck2017feature} estimated fMLLR features in an unsupervised manner. Works in \cite{shibata2017composite,Feng2018exploiting} estimated fMLLR using a pre-trained out-of-domain ASR. Chen et al. \cite{chen2017multilingual} applied vocal tract length normalization (VTLN).
Another direction is to employ DNNs. Zeghidour et al. \cite{Zeghidour+2016} proposed to train subword and speaker same-different tasks within a triamese network and untangle linguistic and speaker information. Chorowski et al. \cite{Chorowski2019unsup} defined a speaker embedding as a condition of VAE decoder to free the encoder from capturing speaker information. Tsuchiya et al. \cite{Tsuchiya2018speaker} applied speaker adversarial training in a  task related to the zero-resource scenario but transcription for a target language was used in model training.

In this paper, we propose to extend our recent research findings \cite{Feng2019improving} on applying disentangled speech representation learned from factorized hierarchical VAE (FHVAE) models \cite{hsu2017nips}  to improve speaker-invariant subword modeling. The contributions made in this study are in several aspects. 
First, the  FHVAE based speaker-invariant learning  is compared with  speaker adversarial training in the strictly unsupervised scenario. 
Second, the combination of adversarial training and disentangled representation learning is studied.  
Third, our proposed approaches are evaluated on the latest challenge ZeroSpeech 2019, as well as on ZeroSpeech 2017 for completeness.
To our best knowledge, direct comparison of the two approaches and their combination has not been studied before.

\section{System description}
\subsection{General framework}
\begin{figure}[t]
    \centering
    \includegraphics[width=1 \linewidth]{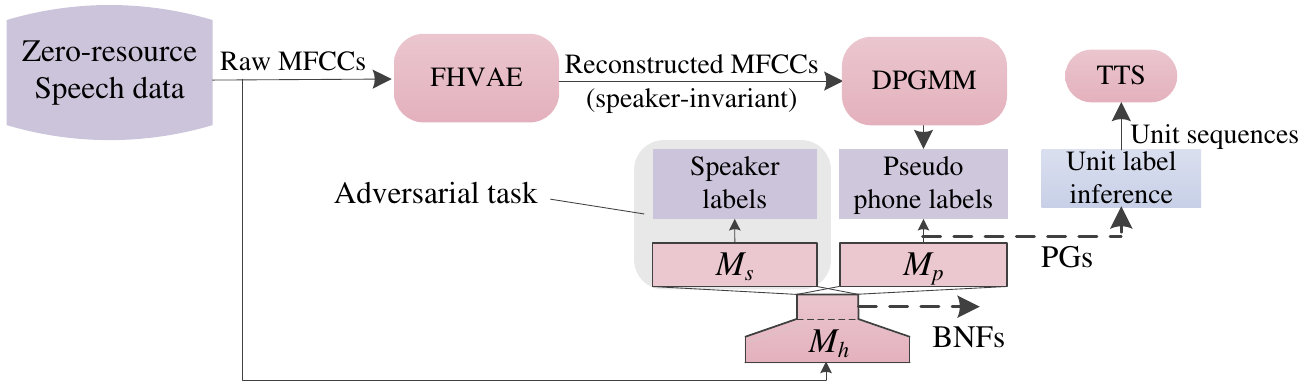}
    \caption{General framework of our proposed approaches}
    \label{fig:framework}
\end{figure}
The general framework of our proposed approaches is illustrated in Figure \ref{fig:framework}. Given untranscribed speech data, the first step is to learn speaker-invariant features   to support frame labeling. The FHVAE model \cite{hsu2017nips} is adopted for this purpose. FHVAEs disentangle linguistic content and speaker information encoded in speech into different latent representations. 
Compared with raw MFCC features, FHVAE reconstructed features conditioned on latent linguistic representation are expected to keep linguistic content unchanged and are more speaker-invariant. 
Details of the FHVAE structure and feature reconstruction methods are described in Section \ref{subsec:fhvae}.

The  reconstructed features are fed as inputs to Dirichlet process Gaussian mixture model (DPGMM) \cite{chang2013parallel} for frame clustering, as was done in \cite{chen2015parallel}. 
The frame-level cluster labels are regarded as pseudo phone   labels to support supervised DNN training. Motivated by successful applications of adversarial training \cite{ganin2015unsupervised} in a wide range of   domain invariant learning tasks \cite{sun2017unsupvised,meng2018speaker,yi2019language,peng2019adversarial}, this work proposes to add an auxiliary adversarial speaker classification task  to explicitly target speaker-invariant feature learning.
After speaker adversarial multi-task learning (AMTL) DNN training, 
softmax PG representation from pseudo phone classification task is used to infer subword unit sequences.
The resultant unit sequences are regarded as pseudo transcriptions for subsequent TTS training.

\subsection{Speaker-invariant feature learning by FHVAEs}
\label{subsec:fhvae}
The FHVAE model formulates the generation process of sequential data by imposing sequence-dependent and sequence-independent priors to different latent variables \cite{hsu2017nips}. It consists of an inference model $\phi$ and a generation model $\theta$.
Let $\mathcal{D}=\{\bm{X^{i}}\}_{i=1}^{M}$ denote a speech dataset with $M$ sequences. 
Each $\bm{X^i}$ contains $N^i$ speech segments $\{\bm{x^{(i,n)}}\}^{N^i}_{n=1}$, where $\bm{x^{(i,n)}}$ is composed of fixed-length consecutive 
frames. The FHVAE model generates a sequence $\bm{X}$ from a random process as follows: 
(1) An \textit{s-vector} $\bm{\mu_2 }$ is drawn from a prior distribution $p_{\theta}(\bm{\mu_2})=\mathcal{N} (\bm{0},\sigma^2_{\mu_2} \bm{I})$;
(2) Latent segment variables $\bm{z_1 ^{n}} $ and latent sequence variables $\bm{z_2^{n}} $ are drawn from $p_{\theta}(\bm{z_1 ^{n}})=\mathcal{N} (\bm{0}, {\sigma^2_{z_1}} \bm{I})$ and  $p_{\theta}(\bm{z_2 ^{n}| \bm{\mu_2}})=\mathcal{N}(\bm{\mu_2}, {\sigma^2_{z_2}} \bm{I} )$ respectively;
(3) Speech segment $\bm{x^{n}}$ is drawn from $p_{\theta}(\bm{x^{n}}|\bm{z_1 ^{n}, \bm{z_2^{n}}})=\mathcal{N}(f_{\mu_x} (\bm{z_1 ^{n}}, \bm{z_2^{n}}), diag(f_{\sigma^2_x} (\bm{z_1 ^{n}}, \bm{z_2^{n}}))$.
Here $\mathcal{N}$ denotes standard normal distribution, $ f_{\mu_x} (\cdot, \cdot)$ and $ f_{\sigma^2_x} (\cdot, \cdot)$ are parameterized by DNNs.
The joint probability for $\bm{X}$ is formulated as,
\begin{equation}
    p_{\theta} (\bm{\mu_2})\prod_{n=1}^{N} p_{\theta} (\bm{z_1^n}) p_{\theta} (\bm{z_2^{n}}|\bm{\mu_2})p_{\theta} (\bm{x^n}|\bm{z_1 ^{n}, \bm{z_2^{n}}}).
\end{equation}

Since the exact posterior inference is intractable,  the FHVAE introduces an inference model $q_{\phi}$
to approximate the true posterior,
\begin{equation}
   q_{\phi} (\bm{\mu_2})\prod_{n=1}^{N}q_{\phi} (\bm{z_2^n}| \bm{x^n}) q_{\phi}(\bm{z_1^n}|\bm{x^n}, \bm{z_2^n}).
   \label{eqt:inference}
\end{equation}
Here  $q_{\phi} (\bm{\mu_2}), q_{\phi} (\bm{z_2^n}| \bm{x^n})$ and $q_{\phi}(\bm{z_1^n}|\bm{x^n}, \bm{z_2^n})$ are all diagonal Gaussian distributions. The mean and variance values of $q_{\phi} (\bm{z_2^n}| \bm{x^n})$ and $q_{\phi}(\bm{z_1^n}|\bm{x^n}, \bm{z_2^n})$ are parameterized by two DNNs. For $q_{\phi} (\bm{\mu_2})$, during FHVAE training, a trainable lookup table containing posterior mean of $\bm{\mu_2}$ for each sequence is updated. During testing, maximum a posteriori (MAP) estimation is used to infer  $\bm{\mu_2}$ for unseen test sequences.
FHVAEs optimize the discriminative segmental variational lower bound which was defined in \cite{hsu2017nips}. It contains a  discriminative objective to prevent $\bm{z_2}$ from being the same for all utterances.

After FHVAE training, $\bm{z_1}$ encodes segment-level factors e.g. linguistic information, while $\bm{z_2}$ encodes sequence-level factors that are relatively consistent within an utterance. By concatenating training utterances of the same speaker into a single sequence for FHVAE training, the learned  
 $\bm{\mu_2}$ is expected to be discriminative to speaker identity.
This work considers applying \textit{s-vector unification} \cite{Feng2019improving} to generate reconstructed feature representation that keeps linguistic content unchanged and is more speaker-invariant than the original representation.
Specifically, a representative speaker with his/her s-vector (denoted as $\bm{\mu_2^*}$) is chosen from the dataset. Next, for each speech segment $\bm{x^{(i,n)}}$ of an arbitrary speaker $i$, its corresponding latent sequence variable $\bm{z_2^{(i,n)}}$ inferred from $\bm{x^{(i,n)}}$ is transformed to $\bm{\hat{z}_2^{(i,n)}}=\bm{z_2^{(i,n)}}-\bm{\mu_2^i}+\bm{\mu_2^*}$, where $\bm{\mu_2^i}$ denotes the s-vector of speaker $i$. 
Finally the FHVAE decoder reconstructs speech segment $\bm{\hat{x}^{(i,n)}}$ conditioned on $\bm{z_1^{(i,n)}}$ and $ \bm{\hat{z}_2^{(i,n)}}$. The features $\{\bm{\hat{x}^{(i,n)}}\}$ form our desired speaker-invariant representation.

\subsection{Speaker AMTL}
\label{subsec:amtl}
Speaker AMTL 
simultaneously trains a subword classification network ($M_p$),
a speaker classification network ($M_s$) and a shared-hidden-layer feature extractor ($M_h$), where $M_p$ and $M_s$ are set on top of $M_h$, as illustrated in Figure \ref{fig:framework}.
In AMTL, the error is reversely propagated from $M_s$ to $M_h$ such that 
the output layer of  $M_h$ is forced to learn   speaker-invariant features so as to confuse $M_s$, while $M_s$ tries to correctly classify outputs of $M_h$ into their corresponding speakers. At the same time, $M_p$ learns to predict the correct DPGMM labels of input features, and back-propagate errors to $M_h$ in a usual way.

Let $\theta_p, \theta_s$ and $\theta_h$ denote the network parameters of
$M_p, M_s$ and $M_h$, respectively.
With the stochastic gradient descent (SGD) algorithm, these parameters are updated as,
\begin{align}
    \theta_p &\leftarrow \theta_p - \delta\frac{\partial \mathcal{L}_{p}}{\partial \theta_p}, 
    \theta_s  \leftarrow \theta_s - \delta \frac{\partial \mathcal{L}_s}{\partial \theta_s}, \\
    \theta_h & \leftarrow \theta_h -\delta \Big[\frac{\partial \mathcal{L}_p}{\partial \theta_h} - \lambda \frac{\partial \mathcal{L}_s}{\partial \theta_h}\Big], \label{eqt:grl}
\end{align}
where $\delta$ is the learning rate, $\lambda$ is the adversarial weight, $\mathcal{L}_p$ and $\mathcal{L}_s$ are the loss values of subword and speaker classification tasks respectively, both in terms of cross-entropy. To implement Eqt. (\ref{eqt:grl}), a gradient reversal layer (GRL) \cite{ganin2015unsupervised} was designed to connect $M_h$ and $M_s$. The GRL acts as identity transform during forward-propagation and changes the sign of loss  during back-propagation. 
After training, the output of $M_h$ is speaker-invariant and subword discriminative bottleneck feature (BNF) representation of input speech. Besides, the softmax output representation of $M_p$ is believed to carry less speaker information than that without performing speaker adversarial training. 
 
\subsection{Subword unit inference and  smoothing}
Subword unit sequences for the concerned
untranscribed speech utterances are inferred from  softmax PG representation of $M_p$ in the speaker AMTL DNN.
For each input frame to the DNN, the DPGMM label with the highest probability in PG representation
is regarded as the subword unit assigned to this frame.  These  frame-level unit labels are further processed by collapsing consecutive repetitive labels to form pseudo transcriptions.

We observed non-smoothness in the inferred unit sequences by using the above methods, i.e.,  frame-level unit labels that are isolated without temporal repetition.
Considering that ground-truth phonemes generally span at least several frames, these non-smooth labels are unwanted. This work proposes an empirical method to filter out part of the non-smooth unit labels, which is summarized in Algorithm \ref{algo}.

\begin{algorithm}[h]
\SetAlgoLined
\KwIn{Frame-level unit labels  $\bm{S}=\{s_1, \ldots ,s_N\}$ }
\KwOut{Pseudo transcription $\bm{T}=\{t_1, \ldots, t_L\}$}
 $\bm{B} \leftarrow \{b_1, \ldots ,b_N$\}, where $b_1 \leftarrow \textit{true}$, $b_j \leftarrow \textrm{bool} (s_j\neq s_{j-1})$ for $2\leq j \leq N$\; 
 \While{$5\leq i\leq N$}{
  \If{$(b_{i-4}=\textrm{true}) and (b_{i-3}=\textrm{true}) and   (b_{i-2}=\textrm{true}) and (b_{i-1} + b_{i} = \textrm{true})  $}{
  $b_{i-4}\leftarrow \textit{false}$; $i \leftarrow i+1$\;
  }
  {}
 }
 $\bm{T} \leftarrow \bm{S} [\textrm{find}(\bm{B[\cdot] = \textit{true}})]$\;
 \caption{Unit sequence smoothing}
 \label{algo}
\end{algorithm}

%

\section{ZeroSpeech 2017 experiments}
\label{sec:exp_setup}
\subsection{Dataset and evaluation metric}
ZeroSpeech 2017 development dataset consists of three languages, i.e. English, French and Mandarin. 
Speaker information  for training sets are given while unknown for test sets. The durations of training sets are $45, 24$ and $2.5$ hours respectively.
Detailed information of the dataset can be found in \cite{dunbar2017zero}.

The evaluation metric is ABX subword discriminability. Basically, it is to decide whether $X$ belongs to $x$ or $y$ if $A$ belongs to $x$ and $B$ belongs to $y$, where $A, B$ and $X$ are speech segments, $x$ and $y$ are two phonemes  that differ in the central sound (e.g., ``beg''-``bag''). Each pair of $A$ and $B$ is spoken by the same speaker. Depending on whether $X$ and $A(B)$ are spoken by the same speaker, ABX error rates for \textit{across-/within-speaker} are evaluated separately. Dynamic time warping and cosine distance are used to measure segment- and frame-level dissimilarity.
\subsection{System setup}
\label{subsec:baseline}
The FHVAE model is trained with merged training sets of all three target languages.
Input features are fixed-length speech segments of $10$ frames. Each frame is represented by a $13$-dimensional MFCC with cepstral mean normalization (CMN) at speaker level. During training, speech utterances spoken by the same speaker are concatenated to a single training sequence. During the inference of hidden variables $\bm{z_1}$ and $\bm{z_2}$, input segments are shifted by $1$ frame. To match the length of latent variables with original features, the first and last frame are padded. 
To generate speaker-invariant reconstructed MFCCs using  the s-vector unification method, a representative speaker  is selected from training sets. In this work the English speaker `s4018' is chosen. The encoder and decoder networks of the FHVAE are both $2$-layer LSTM with $256$ neurons per layer. Latent variable dimensions for $\bm{z_1}$ and $\bm{z_2}$ are $32$. FHVAE  training is implemented in Tensorflow \cite{Abadi2016tensorflow} with the Adam \cite{kingma2014adam} optimizer, using tools developed by \cite{hsu2017nips}. 

The FHVAE based speaker-invariant MFCC features with $\Delta$ and $\Delta\Delta$ are fed as inputs to DPGMM clustering. Training data for the three languages are clustered separately. The numbers of clustering iterations for English, French and Mandarin are $80, 80$ and $1400$. After clustering, the numbers of clusters are $591, 526$ and $314$. Frame  labels are used as pseudo phone alignments to support multilingual DNN training. Speaker labels  obtained from training data support the adversarial speaker classification task. DNN input features are MFCC+CMVN. The layer-wise structure  of   $M_h$ is $\{1024\times 5, 40\}$. Nonlinear function is sigmoid, except the linear BN layer. $M_s$ contains $3$ sub-networks, 
one for each language.
The sub-network contains a GRL, a feed-forward layer (FFL) and a softmax layer. The GRL and FFL are $1024$-dimensional. $M_p$ also contains $3$ sub-networks, each having a $1024$-dimensional FFL and a softmax layer.
During AMTL DNN training, the learning rate starts from $8\cdot 10^{-3}$ to $8\cdot 10^{-4}$ with exponential decay. The number of epochs is $5$. 
Speaker adversarial weight $\lambda$ ranges from $0$ to $0.1$. 
After training, BNFs extracted from $M_h$ are evaluated by the ABX task.
DNN is implemented using Kaldi \cite{povey2011kaldi} \texttt{nnet3} recipe. DPGMM is implemented using tools developed by \cite{chang2013parallel}.

In addition to  DPGMM clustering by FHVAE based speaker-invariant features, we also implement clustering by raw MFCC features and generate alternative DPGMM labels  for comparison. In this case, the numbers of clustering iterations for the three languages are $120,200$ and $3000$.
The numbers of clusters are 
$1118, 1345$ and $596$.
The DNN  structure and training procedure are the same as mentioned above.
\label{subsec:zs2017_setup}
\subsection{Experimental results}

\begin{figure}[t]
    \centering
    \includegraphics[width = \linewidth]{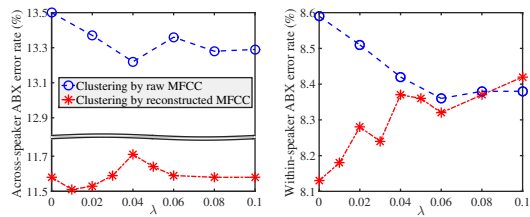}
    \caption{Average ABX error rates on BNF over $3$ languages }
    \label{fig:amtl}
\end{figure}

Average ABX error rates on BNFs over three target languages with different values of $\lambda$ are shown in Figure \ref{fig:amtl}. 
In this Figure, $\lambda=0$ denotes that  speaker adversarial training is not applied. 
From the dashed (blue) lines, it can be observed that speaker adversarial training could reduce ABX error rates in both across- and within-speaker conditions, with absolute reductions of $0.28\%$ and $0.23\%$ respectively. The amount of  improvement is in accordance with the findings reported in \cite{Tsuchiya2018speaker}, despite that  \cite{Tsuchiya2018speaker} exploited English transcriptions during training.
The dash-dotted (red) lines show that when  DPGMM labels generated by reconstructed MFCCs are employed in DNN training, the  positive impact of speaker adversarial training in  across-speaker condition is relatively limited. 
Besides, negative impact is observed in within-speaker condition.
From  Figure \ref{fig:amtl}, it can be concluded  that 
for the purpose of improving the robustness of subword modeling towards speaker variation, frame labeling based on disentangled speech representation learning is more prominent than speaker adversarial training.



\label{subsec:zs2017_results}

\section{ZeroSpeech 2019 experiments}
\subsection{Dataset and evaluation metrics}
ZeroSpeech 2019 \cite{dunbar2019zero} provides untranscribed speech data for two languages. English is used for development while the surprise language (Indonesian) \cite{sakti2008development_ococosda,sakti2008development_tcast} is used for test only. Each language pack consists of training and test sets. The training set consists of a \textbf{unit} discovery dataset for building unsupervised subword models, and a \textbf{voice} dataset for training the TTS system. The optional parallel dataset is not used in this work. Details of ZeroSpeech 2019 datasets are listed in Table \ref{tab:zr19_data}.
\begin{table}[t]
\renewcommand\arraystretch{1}
\centering
\caption{ZeroSpeech 2019 datasets}
\resizebox{0.8 \linewidth}{!}{%
\begin{tabular}{ll|cc|cc}      
 \toprule[1pt]\midrule[0.3pt]

 && \multicolumn{2}{c|}{ English} & \multicolumn{2}{c}{Surprise} \\
&& Duration & \#speakers & Duration & \#speakers \\
\midrule
\multirow{ 2}{*}{Training} & Unit & $15.5$ hrs & $100$ & $15$ hrs &$112$ \\
& Voice & $4.5$ hrs & $2$ & $1.5$ hrs & $1$ \\
\midrule
\multicolumn{2}{c|}{Test} & $0.5$ hr & $24$ & $0.5$ hr & $15$ \\
\midrule[0.3pt]\bottomrule

\end{tabular}%
}
\label{tab:zr19_data}
\end{table}
\begin{figure}[t]
    \centering
    \includegraphics[width=0.85\linewidth]{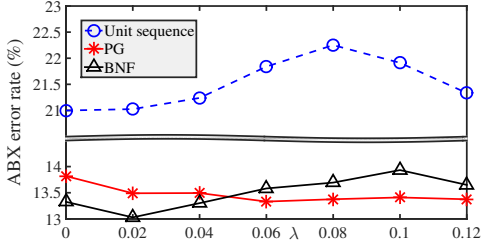}
    \caption{ABX error rates on unit sequence, PG and BNF with different adversarial weights evaluated on English test set}
    \label{fig:adv_results_zs19}
\end{figure}

There are two categories of evaluation metrics in ZeroSpeech 2019.
The metrics for text embeddings, i.e. automatically discovered subword unit sequences  as well as feature representations e.g. BNFs or PGs, are ABX discriminability and bitrate.  ABX error rate is the same as in ZeroSpeech2017. Bitrate is defined as the amount of information provided in the inferred unit sequences.
The metrics for synthesized speech waveforms are character error rate (CER), speaker similarity (SS, $1$ to $5$, larger is better) and mean opinion score (MOS, $1$ to $5$, larger is better), all evaluated by native speakers.

\subsection{System setup}
FHVAE model training and speaker-invariant MFCC reconstruction are performed following the configurations in ZeroSpeech 2017. The unit dataset is used for training. During MFCC reconstruction, a male speaker for each of the two languages is randomly selected as the representative speaker for s-vector unification. Our recent research findings \cite{Feng2019improving} showed that male speakers are more suitable than females in generating speaker-invariant features. The IDs of the selected speakers are  `S015' and `S002' in English and Surprise respectively.
In DPGMM clustering, the numbers of clustering iterations are both $320$. Input features are reconstructed MFCCs+$\Delta$+$\Delta\Delta$. After clustering, the numbers of clusters are $518$ and $693$. 
The speaker AMTL DNN structure and training procedure follow configurations in ZeroSpeech 2017. One difference is the placement of  adversarial sub-network $M_s$. Here $M_s$ is put on top of the FFL in $M_p$ instead of on top of $M_h$.
Besides, the DNN is trained in a monolingual manner. After DNN training, PGs for voice and test sets are extracted. 
BNFs for test set are also extracted. Adversarial weights $\lambda$ ranging from $0$ to $0.12$ with a step size of $0.02$ are evaluated on English test set.

The TTS model is trained with voice dataset and their subword unit sequences inferred from PGs. TTS training is implemented using tools \cite{wu2016merlin}
in the same way as  in the baseline.
The trained TTS synthesizes speech waveforms according to unit sequences inferred from test speech utterances. Algorithm \ref{algo} is applied to voice set and optionally applied to test set.
\subsection{Experimental results}
ABX error rates on subword unit sequences, PGs and BNFs with different values of $\lambda$ evaluated on English test set are shown in Figure \ref{fig:adv_results_zs19}. Algorithm \ref{algo} is not applied at this stage. It is observed that speaker adversarial training could achieve  $0.49\%$ and $0.30\%$ absolute error rate reductions on PG and BNF representations.
The unit sequence representation  does not benefit from adversarial training. Therefore, the optimal $\lambda$ for  unit sequences is $0$.
The  performance gap between frame-level PGs and unit sequences measures the phoneme discriminability distortion caused by the unit inference procedure in this work.

We fix $\lambda=0$ to  train the TTS model, and synthesize test speech waveforms using the trained TTS.
Experimental results of our submission systems are summarized in Table \ref{tab:zr19_results_entire}. 
\begin{table}[t]
\renewcommand\arraystretch{1}
\centering
\caption{Comparison of baseline, topline and our submission}
\resizebox{1 \linewidth}{!}{%
\begin{tabular}{l|ccccc|ccccc}      
 \toprule[1pt]\midrule[0.3pt]
& \multicolumn{5}{c|}{English} & \multicolumn{5}{c}{Surprise}\\
 & ABX & Bitrate& MOS & CER & SS &  ABX & Bitrate& MOS & CER & SS \\
\midrule
 Baseline \cite{dunbar2019zero} & $35.63$ & $71.98$ & $2.50$	 &$0.75$	 & $2.97$ &$27.46$ &$74.55$ &$2.07$&$0.62$	&$3.41$\\
 Topline \cite{dunbar2019zero} &$29.85$ &$37.73$ &$2.77$ &$0.44$ &$2.99$ &$16.09$ &$35.20$ &$3.92$ &$0.28$ &$3.95$ \\
Unit   & ${\color{red}\bm{21.00}}$  &  $413.2$ &$1.56$ & $0.84$&$2.18$ &   ${\color{red}\bm{10.64}}$ & $470.2$ & $1.28$ & $0.74$ & $2.01$\\
Unit+SM & ${\color{red}25.03}$ & $\bm{320.0}$   & $\bm{1.78}$& $\bm{0.76}$&$\bm{2.33}$   &${\color{red}16.87}$ &$\bm{299.2}$ & $\bm{1.67}$ & $\bm{0.66}$ &$\bm{2.60}$\\
\midrule
 PG & $13.82$ & $1733$ & $-$ & $-$ & $-$ &$7.49$ &$1373$&$-$ & $-$ & $-$ \\
 BNF & $13.33$ & $1733$ & $-$ & $-$ & $-$  &$6.52$ & $1373$& $-$ & $-$ & $-$ \\
\midrule[0.3pt]\bottomrule[1pt]
\end{tabular}%
}
\label{tab:zr19_results_entire}
\end{table}
In this Table, `+SM' denotes applying sequence smoothing   towards test set unit labels. Compared with the official baseline, our proposed approaches could significantly improve  unit quality in terms of ABX discriminability. Our system without applying SM achieves $14.6\%$ and $16.8\%$ absolute error rate reductions in English and Surprise sets. If SM is applied, while the ABX error rate increases, improvements in all the other evaluation metrics are observed.
This implies that for the goal of speech synthesis, there is a trade off between quality and quantity of the  learned subword units. Besides, our ABX performance is competitive to, or even better than the supervised topline.

Our systems do not outperform baseline in terms of synthesis quality. One possible explanation is that our learned subword units are much more fine-grained than those in the baseline AUD, making the baseline TTS  less suitable for our AUD system. In the future, we plan to investigate on alternative TTS models to take full advantage of our learned subword units.

\section{Conclusions}
This study tackles robust unsupervised subword modeling in the zero-resource scenario. The robustness  towards speaker variation is achieved by combining speaker adversarial training and FHVAE based disentangled speech representation learning. 
Our proposed approaches are evaluated on  ZeroSpeech 2019 and ZeroSpeech 2017. Experimental results on ZeroSpeech 2017 
show that both approaches are effective  while the latter is more prominent, and that their combination brings further marginal improvement in across-speaker condition.
Results on ZeroSpeech 2019 show that our approaches achieve significant  ABX error rate reduction  to the baseline system. 
The proposed unit sequence smoothing algorithm improves synthesis quality, at a cost of slight decrease in ABX discriminability.



\bibliographystyle{IEEEtran}

\bibliography{mybib}


\end{document}